\documentclass[]{aastex631}
\usepackage{amsmath}
\usepackage{graphicx}
\usepackage{threeparttable}
\usepackage{txfonts}
\usepackage{soul}
\usepackage{bm}
\usepackage{tabularx} 
\usepackage{booktabs}
\usepackage{varwidth}
\usepackage{graphicx}
\usepackage{placeins}

\shorttitle{FRB 20240114A: Periodic Linear Polarization}

\shortauthors{Sun et al.}

\begin{document}
\title{Periodic Modulation of Linear Polarization Fraction in Repeating FRB 20240114A}

\correspondingauthor{Yongfeng Huang}
\email{hyf@nju.edu.cn}

\author[0009-0004-2471-323X]{Fang Sun}
\affiliation{School of Astronomy and Space Science, Nanjing
University, Nanjing 210023, China}

\author[0000-0001-7199-2906]{Yong-Feng Huang}
\affiliation{School of Astronomy and Space Science, Nanjing
University, Nanjing 210023, China}
\affiliation{Key Laboratory of Modern Astronomy and Astrophysics
(Nanjing University), Ministry of Education, Nanjing 210023, China}

\author[0009-0002-8460-1649]{Chen Du}
\affiliation{School of Astronomy and Space Science, Nanjing
University, Nanjing 210023, China}

\author[0000-0002-2191-7286]{Chen Deng}
\affiliation{School of Astronomy and Space Science, Nanjing
University, Nanjing 210023, China}
\affiliation{Key Laboratory of Modern Astronomy and Astrophysics
(Nanjing University), Ministry of Education, Nanjing 210023, China}

\author[0000-0002-3386-7159]{Pei Wang}
\affiliation{National Astronomical Observatories, Chinese Academy of Sciences,
Beijing 100101, China}
\affiliation{Institute for Frontiers in Astronomy and Astrophysics,
Beijing Normal University, Beijing 102206, China}
\affiliation{State Key Laboratory of Radio Astronomy and Technology,
Beijing 100101, China}

\author[0000-0002-9173-4573]{Lun-Hua Shang}
\affiliation{School of Physics and Electronic Science,
Guizhou Normal University, Guiyang 550001, China}
\affiliation{Guizhou Provincial Key Laboratory of Radio Astronomy and Data Processing,
Guizhou Normal University, Guiyang 550001, China}

\author[0000-0002-5238-8997]{Chen-Ran Hu}
\affiliation{School of Astronomy and Space Science, Nanjing
University, Nanjing 210023, China}

\author[0009-0000-0467-0050]{Xiao-Fei Dong}
\affiliation{School of Astronomy and Space Science, Nanjing
University, Nanjing 210023, China}

\author[0000-0001-9648-7295]{Jin-Jun Geng}
\affiliation{Purple Mountain Observatory, Chinese Academy of Sciences,
Nanjing 210023, China}

\author[0000-0002-6189-8307]{Ze-Cheng Zou}
\affiliation{Center for Astrophysics, Guangzhou University, Guangzhou 510006, China}

\author[0000-0001-9227-3716]{Nurimangul Nurmamat}
\affiliation{Guangxi Key Laboratory for Relativistic Astrophysics,
School of Physical Science and Technology, Guangxi University,
Nanning 530004, China}

\author[0000-0003-3230-7587]{Orkash Amat}
\affiliation{School of Astronomy and Space Science, Nanjing
University, Nanjing 210023, China}
\affiliation{State Key Laboratory of Radio Astronomy and Technology, National Astronomical Observatories, CAS, Beijing 100101, China}

\author[0009-0006-2322-4981]{Jing-Peng Lyu}
\affiliation{School of Astronomy and Space Science, Nanjing
University, Nanjing 210023, China}

\author[0009-0003-1008-5220]{Yun-Chuan Zhang}
\affiliation{School of Astronomy and Space Science, Nanjing
University, Nanjing 210023, China}

\begin{abstract}
Long-term polarization changes in repeating fast radio bursts
(FRBs) can probe their emission geometry and propagation
environment. We searched for periodic variation in the degree of
linear polarization (DOL) of FRB~20240114A using the polarization
catalog from the Five-hundred-meter Aperture Spherical radio
Telescope. The sample contains 6,099 bursts detected on 95
observing days over $\sim 488$ days. The daily mean DOL series was
analyzed using phase distance correlation and Lomb--Scargle
periodograms over trial periods of 10--240 days. The strongest
peaks occur at $119.91^{+2.67}_{-2.04}$ and
$122.31^{+2.63}_{-2.82}$ days, respectively. The quoted ranges correspond to the period intervals above
the global $3\sigma$ thresholds. Global
false-alarm probabilities were estimated from \(10^7\) random
permutations of the daily values among the fixed observing dates,
with the full period search repeated for each permutation. The
resulting probabilities are \(9.98\times10^{-4}\) and
\(8.95\times10^{-4}\), corresponding to equivalent Gaussian
significances of \(3.09\sigma\) and \(3.12\sigma\), respectively.
As a supplementary check, the same analysis using daily median DOL
yielded similar peaks near 120 days. A descriptive sinusoidal fit
to all 95 daily mean DOL measurements at a fixed period of 119.91
days has a semi-amplitude of 3.23\%. Possible
explanations include precession-driven changes in viewing geometry
and variations in propagation conditions within a binary system.
\end{abstract}


\section{Introduction}
\label{sec:intro}

Fast radio bursts (FRBs) are bright, millisecond-duration radio
transients \citep{Lorimer2007}. Some sources generate repeated
bursts \citep{Spitler2016}. The second CHIME/FRB catalog contains
3,641 sources, including 83 known repeaters \citep{CHIMEFRB2026}.
Repeated observations of these sources allow changes in their
emission and propagation environments to be investigated within a
single object.

Polarization provides constraints on both the emission geometry
and the intervening magnetized plasma. The diverse polarization
position-angle swings observed in FRB~20180301A favor a
magnetospheric origin for its radio emission \citep{Luo2020}. The
Faraday rotation measure (RM) depends on the electron density and
the magnetic-field component along the propagation path. In
FRB~20121102A, an RM of order \(10^5\ \mathrm{rad\,m^{-2}}\) and
its temporal changes indicate an extreme, evolving magnetized
environment \citep{Michilli2018}. The degree of linear
polarization (DOL) is the fraction of the observed burst intensity
that is linearly polarized. \citet{Feng2022} found decreasing
linear polarization toward lower frequencies in several repeaters
and modeled this behavior through RM scatter. In this picture,
radiation traveling along different paths undergoes different
Faraday rotations, reducing the net linear polarization when the
signals are combined. Tracking DOL over time can therefore help
test changes in the emission geometry or propagation conditions.

FRB~20240114A was discovered by CHIME/FRB on 2024 January 14
\citep{Shin2026}. An observation with the Robert C. Byrd Green
Bank Telescope (GBT) detected 365 bursts in 1.38 hours,
demonstrating the high activity of this source. Among 301 bursts
with measured RMs in the GBT sample, approximately 81\% had DOL
above 90\% \citep{Xie2025}. The Five-hundred-meter Aperture
Spherical radio Telescope (FAST) polarization catalog of
FRB~20240114A provides measurements for more than 6,000 bright
bursts from observations spanning approximately 16 months. It also
reports an RM decline of approximately \(200\
\mathrm{rad\,m^{-2}}\) over 200 days, demonstrating long-term
evolution of the magnetized environment \citep{Wang2026}. This
extensive polarization sample provides a basis for investigating
the long-term behavior of DOL.

Periodic activity has been detected in FRB~20180916B, with a cycle
of 16.35 days \citep{CHIMEFRB2020}. For FRB~20240114A,
\citet{Zhou2026} reported a \(143.40\pm7.19\)-day period in burst
activity. \citet{Li2026} reported a 112.91-day modulation in burst
central frequency and compared FAST polarization distributions
between phase intervals defined by that period. These findings
motivate a direct search for periodicity in DOL. Such a search
allows the preferred modulation timescale and its significance to
be evaluated from the polarization measurements themselves.

In this study, we analyze the long-term periodic modulation in the
DOL of FRB~20240114A using the FAST polarization catalog. The
daily mean DOL series is analyzed using phase distance correlation
and Lomb--Scargle periodograms. Global significance is assessed by
randomly reassigning the daily values among the fixed observing
dates. Daily medians are further used as a supplementary check.
The structure of our paper is organized as follows. Section~2
describes the data, and Section~3 presents the methods. The
results are given in Section~4, followed by a brief discussion on
the physical interpretations in Section~5. Section~6 compares our
results with previous activity and central-frequency periods, and
Section~7 summarizes the study.

\section{Data}
\label{sec:data}

\subsection{FAST Polarization Catalog}
\label{subsec:catalog}

The data analyzed in this work were collected from the publicly
released FAST polarization catalog of FRB~20240114A
\citep{Wang2026}. FAST monitored the source from 2024 January 28
to 2025 May 30 in 97 observing sessions, totaling 57.99~hr. The
observations used the central beam of the FAST 19-beam receiver
\citep{Nan2011} in tracking mode and covered 1.0--1.5~GHz. The
campaign detected 17,356 bursts. Polarization measurements were
released for a subset of these detections. We used the file
\texttt{FAST\_FRB20240114A\_pol\_catalog\_v5.csv}, publicly
available from the Science Data Bank
repository.\footnote{\url{https://doi.org/10.57760/sciencedb.fastro.00040}}
It contains 6,134 burst entries. The burst-level DOL measurements
tabulated by \citet{Wang2026} were used directly in this analysis.

\subsection{Sample Selection and Daily DOL Series}
\label{subsec:daily-series}

All 6,134 entries in this catalog had finite values of topocentric arrival time, $\mathrm{S/N}$, and DOL.
Applying $\mathrm{S/N}>20$ to the tabulated values excluded 35 bursts.
Of these, 27 had $\mathrm{S/N}<20$ and eight had $\mathrm{S/N}=20$.
The final sample contains 6,099 bursts from 95 observing days.
The selected bursts span MJD~60337.238--60824.896, giving a temporal baseline of 487.66~days.

The number of detected bursts varies substantially among observing
days. A periodicity search based on all individual bursts would
therefore give greater weight to days with more detections. The
selected bursts were instead grouped according to the integer part
of their topocentric MJD. For each observing day, the daily DOL
was calculated as the unweighted arithmetic mean of the selected
bursts' DOL measurements. The representative time was defined as
the median arrival time of those bursts. This yielded a daily-mean
DOL series containing 95 measurements. Measurement noise can cause
DOL estimates to exceed 100\% \citep{Wang2026}. Cataloged values
above 100\% were retained without clipping.

\section{Methods}
\label{sec:methods}

In this study, we use two methods to analyze periodicity in the
daily-mean DOL series, the Phase distance correlation (PDC) method
and the generalized Lomb--Scargle (GLS) method. Here we first
describe the two methods one by one.

\subsection{Two Methods of Period Searches}
\label{sec:pdc_gls}

Phase distance correlation (PDC) tests for a general dependence
between DOL and the phase defined by a trial period. The
periodogram is based on distance correlation
\citep{Szekely2007,Zucker2018}. For two daily measurements,
$(t_i,D_i)$ and $(t_j,D_j)$, where $t_i$ is the representative
time of observing day $i$, and $D_i$ is its daily-mean DOL, the
distance between their DOL values is
\begin{equation}
a_{ij}=\left|D_i-D_j\right|.
 \label{eq:pdc_dol_distance}
\end{equation}
For a trial period $P$, the wrapped separation between their
observing times is
\begin{equation}
\phi_{ij}(P)=(t_i-t_j)\bmod P.
\label{eq:pdc_wrapped_separation}
\end{equation}
Here, $\phi_{ij}(P)$ lies in $[0,P)$ and has units of days.
The corresponding phase distance is
\begin{equation}
b_{ij}(P)=\phi_{ij}(P)\left[P-\phi_{ij}(P)\right].
\label{eq:pdc_phase_distance}
\end{equation}

The DOL-distance and phase-distance matrices were then double-centered by subtracting their row and column means and adding their grand means.
Let $A_{ij}$ and $B_{ij}$ denote the centered entries corresponding to $a_{ij}$ and $b_{ij}(P)$, respectively.
The PDC statistic was calculated as their normalized inner product:
\begin{equation}
\mathrm{PDC}(P)=
\frac{\displaystyle\sum_{i,j}A_{ij}B_{ij}}
{\displaystyle
\sqrt{
\left(\sum_{i,j}A_{ij}^{2}\right)
\left(\sum_{i,j}B_{ij}^{2}\right)
}}.
\label{eq:pdc_statistic}
\end{equation}
A larger PDC value indicates a stronger dependence between DOL and phase at the corresponding trial period.
Because the statistic compares pairwise distance structures, it does not require a specified modulation profile.

The generalized Lomb--Scargle periodogram (GLS) searches for sinusoidal variation in unevenly sampled data \citep{Lomb1976,Scargle1982,Zechmeister2009,VanderPlas2018}.
In this generalized formulation, the constant offset is fitted together with the sinusoidal terms.
It is referred to as Lomb--Scargle below.
At each trial frequency $f$, we fitted the daily DOL values with a floating-mean sinusoidal model:
\begin{equation}
D(t)=C+a\cos(2\pi f t)+b\sin(2\pi f t),
\label{eq:gls_model}
\end{equation}
where $C$, $a$, and $b$ were determined by least squares.
All observing days were assigned equal weight.
We defined the Lomb--Scargle power as
\begin{equation}
\mathcal{P}_{\mathrm{GLS}}(f)
=
1-\frac{\mathrm{RSS}(f)}{\mathrm{RSS}_{0}},
\label{eq:gls_power}
\end{equation}
where $\mathrm{RSS}(f)$ is the residual sum of squares from the best-fitting sinusoidal model at trial frequency $f$, and $\mathrm{RSS}_{0}$ is that from the constant model.
A larger Lomb--Scargle power indicates a greater reduction in the residual sum of squares relative to the constant model.

The same frequency grid was used for both methods.
The trial periods ranged from 10 to 240 days.
The grid was uniform in frequency, with a spacing of
\begin{equation}
\Delta f=\frac{1}{\mathrm{OS}\,T},
\label{eq:frequency_spacing}
\end{equation}
where $T$ is the time span of the daily series and $\mathrm{OS}=200$ is the frequency oversampling factor.
This gives a frequency spacing of approximately $1.025\times10^{-5}\,\mathrm{d}^{-1}$ and 9348 trial frequencies.
The peak period was determined separately for each method.

\subsection{Global Significance Assessment}
\label{subsec:global-significance}

We assessed the global significance of each peak using a permutation test.
In each permutation, the daily DOL values were randomly reassigned among the fixed observing dates.
The complete 10--240 day period range was searched again, and the largest PDC statistic or Lomb--Scargle power was retained.
Repeating this procedure produced the null distribution of the maximum statistic over the full search range.
The resulting significance therefore accounts for the search over all 9348 trial frequencies \citep{Frescura2008,Baluev2008}. This permutation test assumes that the daily DOL values are exchangeable among the observing dates under the null hypothesis. Null models with time-correlated variability were not tested.

For each method, $10^{7}$ random permutations were performed.
The 90\%, 95\%, and 99\% global thresholds were obtained from the corresponding quantiles of the permutation maxima.
If $r$ permutation maxima were at least as large as the observed maximum, the global false-alarm probability (FAP) was calculated as
\begin{equation}
\mathrm{FAP}_{\mathrm{global}}
=
\frac{r+1}{M+1},
\label{eq:global_fap}
\end{equation}
where $M=10^{7}$.
Adding one to both the numerator and denominator prevents the estimated FAP from being reported as exactly zero \citep{Phipson2010}. The global FAPs were converted to equivalent Gaussian significances using one-sided tail probabilities.

\section{Results}
\label{sec:results}

\subsection{Periodicity in the Daily Mean DOL Series}
\label{subsec:daily-mean-results}

\begin{figure}[htbp]
\centering
\includegraphics[width=\linewidth]{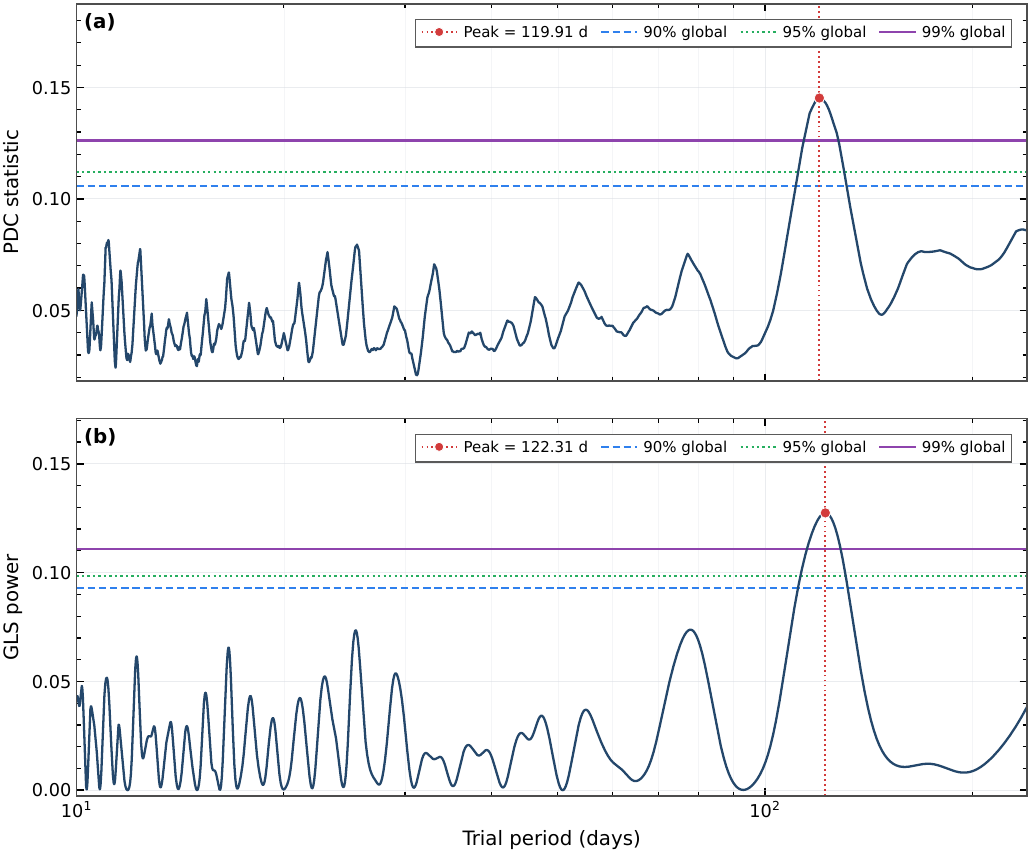}
\caption{Periodograms of the daily mean DOL series of FRB
20240114A. Panels (a) and (b) show the phase distance correlation
statistic and Lomb--Scargle power, respectively, for trial periods
of 10--240 days. Red dotted vertical lines and filled circles mark the maxima at
$119.91^{+2.67}_{-2.04}$ and $122.31^{+2.63}_{-2.82}$ days,
respectively. The quoted ranges correspond to the period
intervals above the global $3\sigma$ thresholds.
Dashed,
dotted, and solid horizontal lines indicate the 90\%, 95\%, and
99\% global thresholds, respectively. These thresholds were
obtained from $10^{7}$ random permutations of the daily DOL values
among the fixed observing dates. The maximum statistic over the
full period range was retained for each permutation. Both observed
maxima exceed their corresponding 99\% global thresholds.}
\label{fig:daily_mean_periodograms}
\end{figure}

Figure~\ref{fig:daily_mean_periodograms} shows the periodograms of the daily mean DOL series. Both methods show their strongest peaks near 120 days. The phase distance correlation periodogram \citep{Zucker2018} reaches its maximum value of 0.1452 at 119.91 days. This value exceeds the 99\% global threshold of 0.1262. The Lomb--Scargle periodogram \citep{Zechmeister2009} reaches its maximum power of 0.1275 at 122.31 days, exceeding its 99\% global threshold of 0.1108.
The period ranges above the global $3\sigma$ thresholds are
$119.91^{+2.67}_{-2.04}$ and $122.31^{+2.63}_{-2.82}$ days
for phase distance correlation and Lomb--Scargle, respectively.

\begin{figure}[htbp]
\centering
\includegraphics[width=\linewidth]{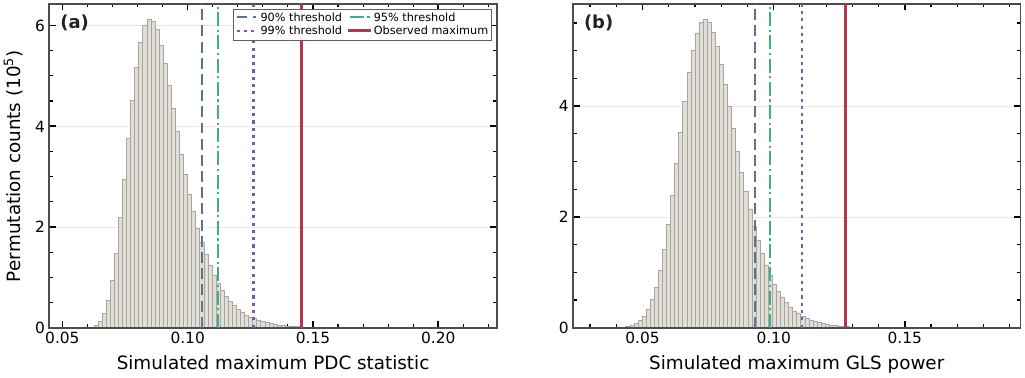}
\caption{Global null distributions of the periodogram maxima for
the daily mean DOL series of FRB 20240114A. Panels (a) and (b)
correspond to phase distance correlation and Lomb--Scargle,
respectively. For each method, $10^{7}$ random permutations were
performed by shuffling the daily DOL values among the fixed
observing dates. The full 10--240-day period range was searched in
each permutation, and the maximum statistic was retained. Gray
histograms show the distributions of these maxima. Dashed,
dash-dotted, and dotted vertical lines mark the 90\%, 95\%, and
99\% global thresholds, respectively. Solid lines mark the maxima
from the observed data. The corresponding global false-alarm
probabilities are $9.98\times10^{-4}$ and $8.95\times10^{-4}$ for
panels (a) and (b), respectively. Both observed maxima exceed
their corresponding 99\% global thresholds.}
\label{fig:daily_mean_global_null_distributions}
\end{figure}

Figure~\ref{fig:daily_mean_global_null_distributions} compares the observed maxima with their global null distributions. Using the estimator of \citet{Phipson2010}, the global FAPs are $9.98\times10^{-4}$ for phase distance correlation and $8.95\times10^{-4}$ for Lomb--Scargle. These correspond to equivalent Gaussian significances of $3.09\sigma$ and $3.12\sigma$, respectively. Both periodograms therefore show a significant peak near 120 days in the daily mean DOL series.

\citet{Zhou2026} reported a period of \(143.40 \pm 7.19\) days in
the burst activity of FRB~20240114A. \citet{Li2026} reported a
112.91-day periodic modulation in its burst central frequency.
These studies probe different observables from the DOL examined
here. A detailed comparison between these reported periods and our
approximately 120-day DOL modulation is presented in Section~6.

\subsection{Supplementary Check Using Daily Median DOL}
\label{subsec:daily-median-check}

\begin{figure}[htbp]
\centering
\includegraphics[width=\linewidth]{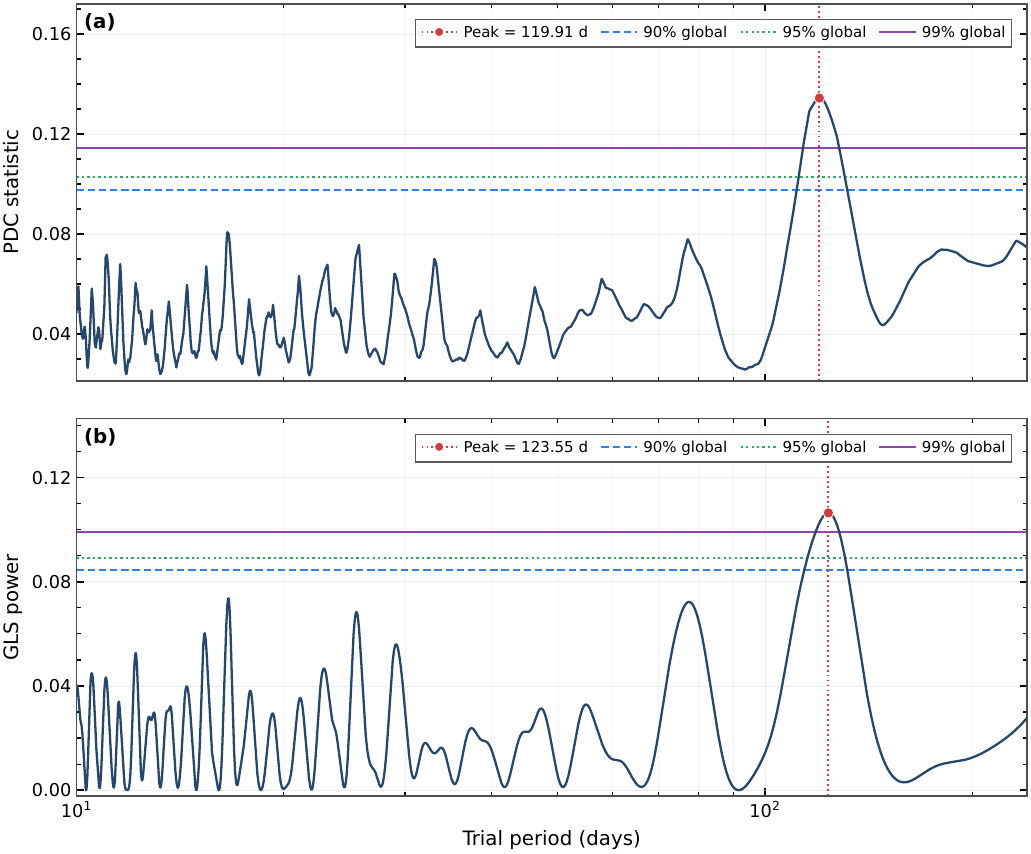}
\caption{Periodograms of the daily median DOL series of FRB
20240114A, used as a supplementary check of the daily mean
results. Panels (a) and (b) show the phase distance correlation
statistic and Lomb--Scargle power, respectively. The maxima at $119.91^{+4.02}_{-4.14}$ and 123.55 days both
exceed their corresponding 99\% global thresholds. The quoted
range corresponds to the period interval above the global
$3\sigma$ threshold for phase distance correlation.
 The plotting conventions are the same as for the daily
mean periodograms in Figure~\ref{fig:daily_mean_periodograms}.}
\label{fig:daily_median_periodograms}
\end{figure}

We have also calculated the median of the selected burst DOL
values on each observing day. Here the median means there are
equal numbers of bursts with DOL larger than and less than that
particular DOL value, which is also a characteristic DOL for the
bursts on that day. As a supplementary check, the same analysis
was repeated using the daily median of DOL values. The search
range and permutation procedure are the same as for the daily mean
periodograms. Figure~\ref{fig:daily_median_periodograms} shows the
resulting periodograms. The strongest peaks occur at 119.91 days
for phase distance correlation and 123.55 days for Lomb--Scargle.
These two periods are well consistent with the daily mean DOL
results.
For phase distance correlation, the period range above the global
$3\sigma$ threshold is $119.91^{+4.02}_{-4.14}$ days.

\begin{figure}[htbp]
\centering
\includegraphics[width=\linewidth]{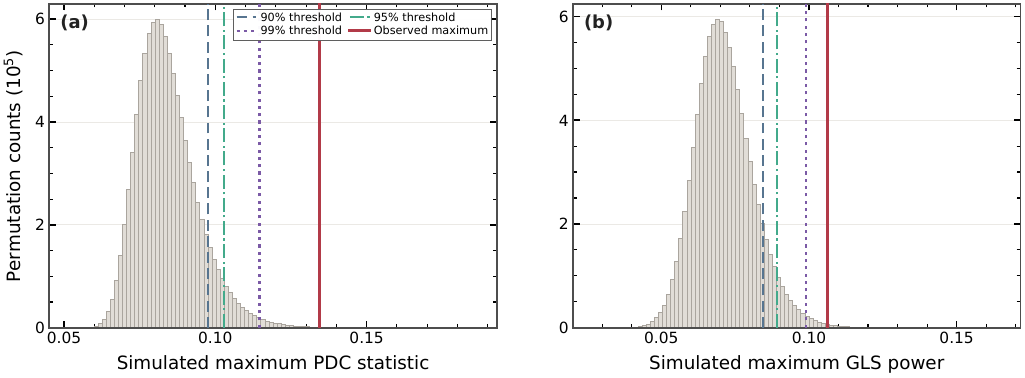}
\caption{Global null distributions of the periodogram maxima for the daily median DOL series of FRB 20240114A, used as a supplementary check of the daily mean results. The panel arrangement, permutation procedure, and line conventions are the same as for the daily mean analysis in Figure~\ref{fig:daily_mean_global_null_distributions}. The global false-alarm probabilities are $5.71\times10^{-4}$ and $2.84\times10^{-3}$ for panels (a) and (b), respectively. Both observed maxima exceed their corresponding 99\% global thresholds.}
\label{fig:daily_median_global_null_distributions}
\end{figure}

Figure~\ref{fig:daily_median_global_null_distributions} compares
the observed maxima with their global null distributions. Both
maxima exceed their corresponding 99\% global thresholds. The
global FAPs are $5.71\times10^{-4}$ for phase distance correlation
and $2.84\times10^{-3}$ for Lomb--Scargle. These correspond to
equivalent Gaussian significances of $3.25\sigma$ and
$2.77\sigma$, respectively. In short, we found that the daily
median analysis also yields peaks near 120 days, consistent with
the daily mean results.

\subsection{Phase-folded DOL and Evolution across Consecutive Cycles}
\label{sec:phase_evolution}

\begin{figure}[htbp]
\centering
\includegraphics[width=\linewidth]{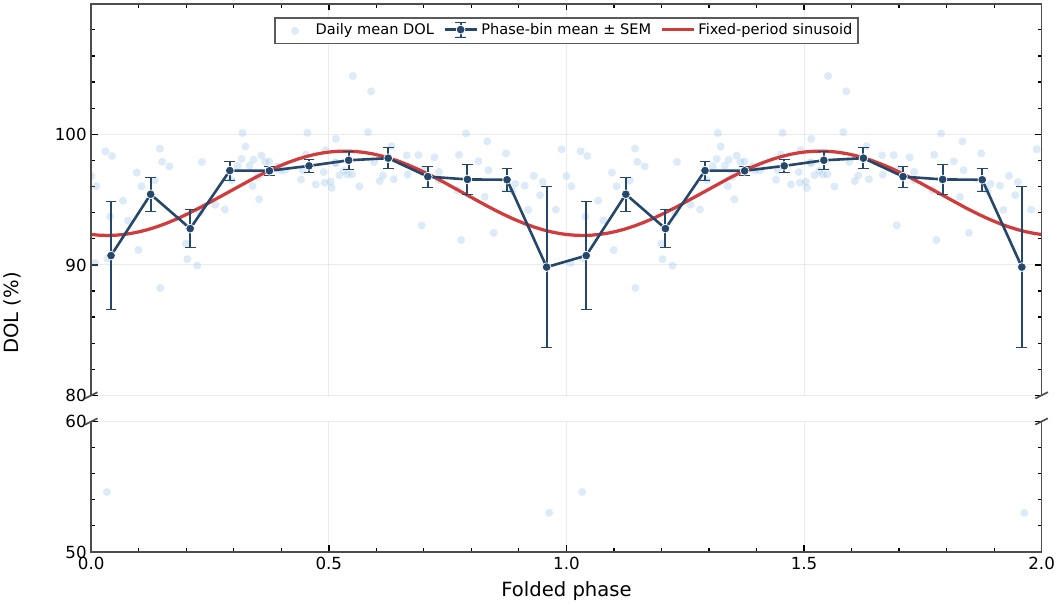}
\caption{Daily mean DOL measurements of FRB 20240114A folded at a fixed period of 119.91 days. Light-blue circles show the 95 daily mean measurements. Dark-blue points show the averages of these daily values in 12 equal-width phase bins. Error bars give the standard error of the mean in each bin. The red curve shows the fixed-period sinusoidal fit to the unbinned daily mean measurements. The measurements, binned profile, and fitted curve are repeated over a second phase cycle for visual continuity.}
\label{fig:dol_phase_folded}
\end{figure}

We have folded the timing series constructed in
Section~\ref{subsec:daily-series} by referring to the 119.91-day
period identified through the phase distance correlation method in
Section~\ref{subsec:daily-mean-results}. The zero phase was set to
the representative time of the first observing day. The phase of
each measurement was calculated as
\begin{equation}
\phi_i
=
\operatorname{mod}\left(\frac{t_i-t_0}{P},1\right),
\label{eq:dol_folded_phase}
\end{equation}
where $t_i$ is the representative time of observing day $i$,
$P=119.91$ days, and $t_0$ is the reference time.
Figure~\ref{fig:dol_phase_folded} shows the phase-folded daily
mean DOL measurements.

The unbinned daily mean measurements were then fitted with the
sinusoidal model
\begin{equation}
\mathrm{DOL}(\phi)
=
C+a\cos(2\pi\phi)+b\sin(2\pi\phi).
\label{eq:dol_sinusoidal_model}
\end{equation}
The coefficients were determined by ordinary least squares. Each
observing day was assigned equal weight, and the period was held
fixed. As shown in Figure~\ref{fig:dol_phase_folded}, the fit to
all 95 daily measurements has a semi-amplitude of 3.23\%. The fitted DOL is highest near the middle of the phase
cycle and lowest near the phase boundary. For display, the daily
measurements were averaged in 12 equal-width phase bins. The error
bars give the standard error of the mean in each bin.

\begin{figure}[htbp]
\centering
\includegraphics[width=\linewidth]{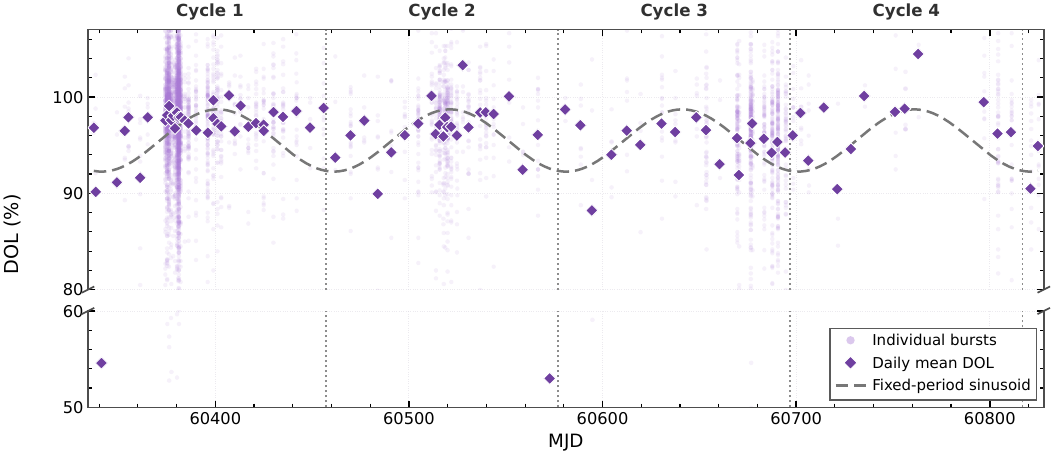}
\caption{DOL evolution of FRB 20240114A over the full observing span of 487.66 days. This span covers approximately 4.07 cycles of the adopted 119.91-day period. Light-purple circles represent individual bursts with ${\rm S/N}>20$, and dark-purple diamonds represent the daily mean DOL values. The gray dashed curve is a single sinusoid fitted to all daily means, with the period fixed at the phase distance correlation peak. The same fit is shown throughout the observing span without refitting individual cycles. Vertical dotted lines mark the cycle boundaries.}
\label{fig:dol_consecutive_cycles}
\end{figure}

Figure~\ref{fig:dol_consecutive_cycles} shows the individual-burst
and daily mean DOL measurements over the full observing span of
487.66 days. It corresponds to approximately 4.07 cycles of the
adopted period. The same fitted sinusoid is shown throughout the
observing span, without refitting individual cycles. The
measurements show substantial scatter around the fitted curve.

\section{Possible Physical Interpretations}
\label{sec:physical}

The $\sim 120$-day DOL modulation could be due to various
mechanisms, including neutron-star free precession, binary-orbital
propagation effects, and other long-term dynamical processes. Here
we present some discussion on these potential explanations.

\subsection{Neutron-star Free Precession}
\label{sec:free_precession}

Free precession has previously been proposed as the clock
underlying periodic activity windows in repeating FRBs
\citep{Levin2020,Zanazzi2020,Li2021}. It could also be one
possible origin of the approximately 120-day DOL modulation in FRB
20240114A. A viable interpretation of the present result requires
two links. First, the precession must satisfy the observed
timescale. Second, the changing viewing geometry must produce a
variation in DOL.

For an approximately biaxial precessing component, let $\mathcal{I}_1$, $\mathcal{I}_2$, and $\mathcal{I}_3$ denote the principal moments of its effective inertia tensor, with $\mathcal{I}_1=\mathcal{I}_2$. The effective ellipticity is
\begin{equation}
\epsilon_{\rm eff}
=
\frac{\mathcal{I}_3-\mathcal{I}_1}{\mathcal{I}_1}.
\label{eq:effective_ellipticity}
\end{equation}
Here the inertia difference represents the stress-supported deformation fixed to the stellar body. The centrifugal bulge that follows the instantaneous spin axis is excluded from this deformation \citep{Jones2001}. In this approximation, the free-precession period is
\begin{equation}
P_{\rm prec}
\simeq
\frac{P_{\rm spin}}
{\left|\epsilon_{\rm eff}\cos\theta\right|},
\label{eq:free_precession_period}
\end{equation}
where $P_{\rm spin}$ is the spin period of the compact central
engine and $\theta$ is the wobble angle between the symmetry axis
and the angular-momentum vector
\citep{Jones2001,Zanazzi2020,Desvignes2024}. If the 119.91-day
phase distance correlation peak is identified with $P_{\rm prec}$,
then
\begin{equation}
\left|\epsilon_{\rm eff}\cos\theta\right|
\simeq
9.65\times10^{-8}
\left(\frac{P_{\rm spin}}{1\ {\rm s}}\right)
\left(\frac{P_{\rm prec}}{119.91\ {\rm days}}\right)^{-1}.
\label{eq:required_ellipticity}
\end{equation}

For spin periods of 1--10~s and small wobble angles, $|\cos\theta|\simeq1$, the required effective ellipticity is approximately $9.65\times10^{-8}$--$9.65\times10^{-7}$. As an observational comparison, the preferred free-precession model of XTE~J1810$-$197 gives a late-time ellipticity of approximately $1.25\times10^{-7}$ \citep{Desvignes2024}. This value falls within the required range. Comparison with whole-star deformation models assumes that most of the stellar moment of inertia participates in the precession \citep{Cutler2003}.

For spin periods of 1--10~ms, the required $|\epsilon_{\rm eff}\cos\theta|$ is approximately $9.65\times10^{-11}$--$9.65\times10^{-10}$. At small wobble angles, these values are the required effective ellipticities. A rotating-star comparison follows from the relaxed-crust calculations of \citet{Cutler2003}. Consider a star with a mass of $1.4\,M_\odot$ and a radius of 12~km. Its crust is assumed to be relaxed at the current spin rate. Elastic stresses supply the body-fixed deformation, and nearly the whole stellar moment of inertia participates in precession. Combining their crustal rigidity and rotational-deformation estimates gives $\epsilon_{\rm eff}\simeq6\times10^{-10}(P_{\rm spin}/10\ {\rm ms})^{-2}$. Under these assumptions, this estimate meets the 119.91-day requirement at a spin period of approximately 8.5~ms, with $\epsilon_{\rm eff}\sim8\times10^{-10}$. Thus, millisecond rotation can accommodate the observed timescale in this specific crust and coupling configuration. Faster spins or a smaller participating moment of inertia shorten the precession period within this scaling.

During free precession, the angular-momentum direction remains approximately fixed, while the body-fixed symmetry and magnetic axes change their orientation relative to the observer. The line of sight can therefore cross different parts of the radio-emission beam \citep{Jones2001,Zanazzi2020,Li2021}. DOL modulation can arise if the sampled regions have different polarization properties. This can occur if the DOL distribution of individual bursts depends on the sampled beam region. It can also occur if the geometry changes the relative detection rates of burst populations with different DOL distributions.

Observations of FRB~20240114A show that its polarization state is complex or time-dependent. The FAST catalog contains a broad, non-uniform distribution of intrinsic polarization position angle (PA) values \citep{Wang2026}. Independent observations also show rapid and stochastic PA variations between and within bursts \citep{Manaswini2026}.

A change in the relative intensities of polarization modes that are unresolved within a Stokes measurement is one possible response to the changing viewing geometry. If two such modes are fully linearly polarized, strictly orthogonal, and mutually incoherent, their combined linear polarization fraction is
\begin{equation}
p_{\rm L}
=
\frac{|I_a-I_b|}{I_a+I_b},
\label{eq:orthogonal_mode_dol}
\end{equation}
where $I_a$ and $I_b$ are the mode intensities \citep{McKinnon1998}. Here $I,Q,U$ are the Stokes parameters of the combined signal, and $p_{\rm L}=\sqrt{Q^2+U^2}/I$. Thus $100p_{\rm L}$ is the DOL expressed as a percentage in this idealized model. Taking $I_b\leq I_a$, the weaker mode contributes a fraction
\begin{equation}
f_{\rm weak}
=
\frac{I_b}{I_a+I_b}
=
\frac{1-p_{\rm L}}{2}
\label{eq:weaker_mode_fraction}
\end{equation}
of the total intensity. The descriptive sinusoid fitted to all 95 daily mean measurements spans approximately $92.25\%$--$98.71\%$ in DOL. As an amplitude illustration, representing these endpoints with the idealized model gives weaker-mode fractions of approximately $3.88\%$ and $0.64\%$, respectively. A small change in mode balance could therefore produce a DOL variation comparable to the range represented by the descriptive fit.

Orthogonal PA jumps have been observed in FRB~20201124A \citep{Niu2024}. For FRB~20240114A, the PA-jump distribution is broad, and jumps near $\pm90^\circ$ are not more common than jumps of other magnitudes \citep{Manaswini2026}. A precession interpretation based on the two-mode response requires mode weights that vary systematically with viewing phase.

The period relation above assumes an approximately biaxial effective precessing component. A real neutron star contains an elastic crust coupled to superfluid components in the inner crust and core. Crustal strain, internal coupling, and vortex pinning can modify the effective deformation and precession dynamics \citep{Jones2001,Cutler2003,Glampedakis2010}. In this simple framework, long-period solutions require weak effective pinning or efficient vortex unpinning. Hydrodynamic forces generated during precession provide a possible unpinning mechanism \citep{Link2002}.

The month-scale damping inferred for XTE~J1810$-$197 highlights the possible evolution of precession after excitation \citep{Desvignes2024}. For FRB~20240114A, the 487.66-day observing span covers approximately 4.07 cycles. The expected evolution of the viewing geometry and DOL amplitude depends on the excitation and damping history.

Phase-dependent changes in the distributions of PA, circular polarization, and burst morphology would provide tests of the geometric interpretation. Changes in the relative occurrence rates of distinct burst populations would provide another test. Simultaneous wideband polarimetry would help separate geometric changes from frequency-dependent propagation effects. A spin-period measurement would constrain the relation between effective ellipticity and wobble angle through the period equation above.

Free precession therefore provides a possible connection between the approximately 120-day timescale and the DOL modulation when the changing line of sight samples emission regions or burst populations with different polarization properties.

\subsection{Binary-orbital Propagation Effects}
\label{sec:binary_propagation}

Orbital motion in a binary system could make the propagation
conditions of FRB~20240114A vary on a long timescale. Binary
models have been proposed for periodic activity in repeating FRBs
\citep{Ioka2020}. A companion wind or circumstellar disk can also
change the RM, scattering, and polarization of radio bursts along
the orbit \citep{Wang2022,Yang2023,Zhao2023}. The FAST catalog
shows that the RM of FRB~20240114A evolves with time
\citep{Wang2026}. These RM variations indicate a changing
magneto-ionic environment along the line of sight.

If the DOL period of 119.91 days is associated with an orbital
period, the Kepler's third law gives the semimajor axis of the
relative orbit:
\begin{equation}
a=
\left(\frac{GM_{\rm tot}P_{\rm orb}^{2}}{4\pi^{2}}\right)^{1/3}
\simeq
1.03\,{\rm au}
\left(\frac{M_{\rm tot}}{10\,M_\odot}\right)^{1/3}
\left(\frac{P_{\rm orb}}{119.91\,{\rm d}}\right)^{2/3}.
\label{eq:binary_separation}
\end{equation}
Here $M_{\rm tot}$ is the total mass of the binary. The
normalization $M_{\rm tot}=10\,M_\odot$ is adopted for
illustration. DOL modulation requires the orbital motion to change either
the DOL of individual bursts or the relative proportions of
bursts with different DOL values in the detected sample.

One direct propagation effect is external Faraday dispersion. A
uniform RM rotates the polarization position angle but does not
reduce the linear polarization fraction when the rotation is
adequately resolved and corrected. Depolarization instead occurs
if unresolved propagation paths undergo different amounts of
Faraday rotation. For a screen with a Gaussian RM distribution,
the Burn relation is \citep{Burn1966,Feng2022}
\begin{equation}
p_{\rm L}(\lambda,\psi)
=
p_{{\rm L},0}
\exp\!\left[
-2\sigma_{\rm RM}^{2}(\psi)\lambda^{4}
\right],
\label{eq:binary_burn}
\end{equation}
where $\lambda$ is wavelength, $\psi$ is orbital phase, and $\sigma_{\rm RM}$ is the unresolved RM dispersion. In this propagation-only model, the intrinsic fraction $p_{{\rm L},0}$ is taken to be independent of orbital phase. An orbit could then modulate DOL if it changes the RM dispersion along the line of sight.

\citet{Wang2026} tested frequency-dependent depolarization in 18 bright, broadband bursts from FRB~20240114A. They found no systematic decrease in DOL toward lower frequencies and obtained a 95\% upper limit of $\sigma_{\rm RM}<0.51\,{\rm rad\,m^{-2}}$ for those bursts. Substituting this limit into the Burn relation gives a maximum reduction of approximately 0.42\% at 1.0~GHz and 0.083\% at 1.5~GHz, assuming an intrinsic linear polarization fraction of 100\%. If the upper limit applies throughout the long-term cycle, these values also bound the maximum phase-to-phase DOL variation produced by the screen. They are well below the 6.46\% peak-to-peak variation of the descriptive sinusoid fitted to the daily mean DOL series. Under this assumption, a single external screen obeying this relation would be insufficient to reproduce the full fitted variation. This comparison is conditional because the published limit constrains the selected bursts, rather than every phase of the long-term cycle.

Other effects within a binary environment need not follow the Burn relation. If different polarization modes experience different optical depths near a cyclotron resonance, changes in their relative transmission could alter the observed DOL \citep{Li2026}. Alternatively, phase-dependent absorption, scattering, or plasma lensing could change which bursts enter the detected sample \citep{Cordes2017,Main2018}. This would change the daily DOL statistics only if the affected burst populations have different DOL distributions.

Broadband polarization measurements covering the full long-term phase could test these alternatives. External Faraday dispersion predicts stronger low-frequency depolarization when DOL is lower. Sample selection instead requires changes in the properties and relative numbers of the detected bursts. An orbital explanation of the DOL modulation therefore requires propagation or sample-selection effects that recur with orbital phase.

\subsection{Other Possible Origins of the Long-term Modulation}

A torque acting on the neutron star could cause the direction of its angular momentum to precess. If $\boldsymbol{L}$ is the angular momentum and $\boldsymbol{N}$ is the torque, $\mathrm{d}\boldsymbol{L}/\mathrm{d}t=\boldsymbol{N}$. A component of $\boldsymbol{N}$ perpendicular to $\boldsymbol{L}$ changes the direction of $\boldsymbol{L}$. Torques from a fallback disk and anomalous electromagnetic torques have both been proposed as drivers of forced precession in repeating FRBs \citep{Tong2020,Sobyanin2020}. If the emission has angularly dependent polarization, the changing viewing geometry could alter the relative contributions of polarized emission components and thereby modulate DOL. The precession period would depend on the strength and geometry of the torque.

Precession associated with a tilted disk provides another possible clock. In the weak-field limit, the local Lense--Thirring precession angular frequency at radius $r$ around a spinning compact object is
\begin{equation}
\Omega_{\rm LT}(r)=\frac{2GJ}{c^{2}r^{3}},
\label{eq:lt_frequency}
\end{equation}
where $J$ is the magnitude of the compact object's angular momentum. The period of an extended disk also depends on its radial structure. \citet{Chen2020} proposed a model in which a tilted debris disk drives precession of the neutron star and inner disk. Precession of the polar funnel in a super-Eddington accreting binary has also been proposed to yield periods from days to years \citep{Sridhar2021}. To produce the observed DOL variation, such precession would have to change either the polarized emission visible to the observer or the propagation conditions along the line of sight.

Another possibility is that the approximately 120-day timescale is
the neutron star's rotation period itself. Particle winds,
flare-induced angular-momentum loss, and fallback disks have been
investigated as ways to produce ultra-long-period magnetars
\citep{Beniamini2020}. Disk-wind calculations also permit periods
of tens to hundreds of days, but only under particularly strong
angular-momentum extraction \citep{Xu2021}. In this
interpretation, rotation could modulate DOL if different
rotational phases expose emission with different polarization
properties.

The energy budget places a further constraint on a rotation-powered version of this model. Taking a stellar moment of inertia $I=10^{45}\,\mathrm{g\,cm^{2}}$ and $P_{\rm spin}=120\,\mathrm{d}$, the rotational energy is
\begin{equation}
E_{\rm rot}
=
\frac{2\pi^{2}I}{P_{\rm spin}^{2}}
\simeq
1.84\times10^{32}
\left(\frac{I}{10^{45}\,\mathrm{g\,cm^{2}}}\right)
\left(\frac{P_{\rm spin}}{120\,\mathrm{d}}\right)^{-2}
\,\mathrm{erg}.
\label{eq:ultralong_spin_energy}
\end{equation}
This is far below the reported isotropic-equivalent energies of individual bursts from this source, approximately $10^{37}$--$10^{39}\,\mathrm{erg}$ \citep{Wang2025}. Isotropic-equivalent energy is not the true emitted energy if the radiation is beamed. Nevertheless, the large difference disfavors the star's current rotational energy as the main burst-energy reservoir unless the radio emission is extremely narrowly beamed and the conversion of rotational energy into radio emission is sufficiently efficient. Magnetic energy could power the bursts while rotation sets the phase of the DOL variation.

These mechanisms assign the approximately 120-day timescale to different physical clocks: forced precession of the neutron star, precession associated with a disk, or stellar rotation. Each also requires the observed polarization to change with phase in order to produce the DOL variation.

\section{Discussion}
\label{sec:discussion}

\citet{Zhou2026} reported a \(143.40 \pm 7.19\)-day period in the burst activity of FRB~20240114A. Their analysis used 111 FAST observations spanning 612.33 days and incorporated weights proportional to the burst rate. Our DOL analysis spans 487.66 days and assigns equal weight to each observing day. Over this 487.66-day interval, the 119.91-day DOL model completes 4.07 cycles, whereas a 143.40-day model completes 3.40 cycles. The relative phase of these two fixed-period models would therefore drift by approximately 0.67 cycles. If the activity period represents an orbital period and the propagation conditions repeat each orbit, an associated DOL variation should also repeat with orbital phase. Activity and DOL maxima could occur at different phases, but a constant phase offset would not explain the difference between the fitted periods. A shared orbital interpretation would therefore need to reproduce both datasets with a common period, while allowing their modulation profiles and phases to differ.

An activity cycle can reflect changes in burst production, visibility, or both. If only the burst rate varies while the polarization distribution remains unchanged, the expected daily mean DOL also remains unchanged. A geometric connection is possible if changing viewing directions both alter detectability and sample emission regions with different polarization fractions \citep{Li2021}. In a binary environment, the companion's magnetized plasma could affect both visibility and polarization through absorption and polarization transfer \citep{Xia2023}. The resulting DOL would depend on how propagation modifies the polarization of the detected emission and which burst populations remain detectable. The sign and phase of an activity--DOL association would depend on these physical responses. An activity maximum would therefore not necessarily coincide with a DOL maximum.

A separate question is whether DOL and burst central frequency vary on comparable long timescales. \citet{Li2026} reported a 112.91-day modulation in individual Parkes burst central frequencies using the Lomb--Scargle method. Their phase-folding tests with 5--20 phase bins gave a period
range of 104--123 days. For our comparison, a combined daily-median central-frequency series was constructed from Parkes \citep{Uttarkar2026}, Effelsberg \citep{Limaye2026}, and GBT measurements \citep{Xie2025}. The series contains 110 daily values. All entries with both an observing time and a measured central frequency were included, without additional burst-level cuts. Bursts observed on the same day were pooled, and the median central frequency and median burst MJD were adopted for that day. An equal-day-weight, floating-mean Lomb--Scargle search was performed over periods from 2 days to one-third of the total time span \(T\), with a frequency spacing of \(1/(200T)\). The strongest peak occurs at 112.90 days, which was adopted for the central-frequency sinusoidal fit.

\begin{figure*}[t]
\centering
\includegraphics[width=\textwidth]{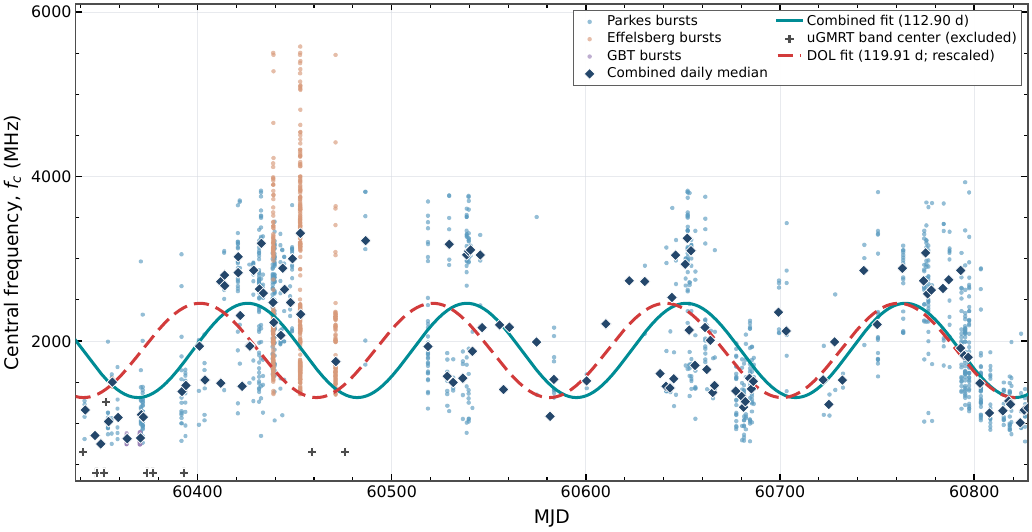}
\caption{Comparison of the DOL and central-frequency sinusoids of
FRB 20240114A on the same MJD axis. Blue, coral-orange, and
purple circles show individual-burst central frequencies
measured with Parkes \citep{Uttarkar2026}, Effelsberg
\citep{Limaye2026}, and GBT \citep{Xie2025}, respectively.
Dark-blue diamonds show the daily medians of the combined
three-telescope sample. The teal solid curve is a sinusoidal fit
to this series at a period of 112.90 days. The red dashed curve
shows the 119.91-day sinusoidal fit to the daily mean DOL series.
The DOL curve is shifted and scaled vertically to match the fitted
mean and semi-amplitude of the central-frequency curve. This
rescaling preserves its peak times for comparison. Gray crosses
mark the uGMRT receiver-band centers at the observing epochs
\citep{Kumar2024}, which are shown for reference but are excluded
from the central-frequency fit.} \label{fig:dol_combined}
\end{figure*}

Figure~\ref{fig:dol_combined} compares this fit with the 119.91-day sinusoid fitted to all 95 daily mean DOL measurements. The DOL curve is shifted and scaled vertically to match the fitted mean and semi-amplitude of the central-frequency curve. This rescaling preserves its peak times. The periods and phases of the two curves were determined separately. During the earlier cycles, the DOL peaks precede the central-frequency peaks. The peak times become closer later in the observing span. The different fitted periods produce a gradual change in their relative peak times, so the apparent alignment varies across the observations.

The local similarities between the DOL and central-frequency trends can be considered within the geometric and propagation scenarios discussed above. Different viewing directions may sample regions with different emission frequencies and linear polarization fractions. A changing viewing geometry could then affect both observables \citep{Zanazzi2020}. Frequency-dependent propagation could reshape the spectrum and change the relative weights of frequency components with different polarization properties. This would also change the band-integrated DOL \citep{Cordes2017,Feng2022}. A change in the detected burst population could affect both quantities if the selected populations differ in their central-frequency and polarization distributions \citep{Uttarkar2026}.

Differences in telescope bandpasses, sensitivities, and observing cadence affect which bursts enter the combined central-frequency series. The DOL and central-frequency measurements also generally come from different bursts. Figure~\ref{fig:dol_combined} therefore provides a qualitative comparison of their long-term timing. The local similarities motivate considering a possible connection, while the changing peak offsets and differences between the samples must be accounted for in any interpretation of a shared modulation.

\section{Summary}
\label{sec:summary}

We analyzed the FAST polarization catalog of FRB~20240114A for
long-term variation in DOL. The sample contains 6,099 bursts
detected on 95 observing days over 487.66 days. The daily mean
DOL series was analyzed using phase distance correlation and
Lomb--Scargle periodograms. The strongest peaks occur at $119.91^{+2.67}_{-2.04}$ and
$122.31^{+2.63}_{-2.82}$ days, respectively. 
The quoted ranges
correspond to the period intervals above the global $3\sigma$
thresholds. Their global false-alarm
probabilities, estimated from \(10^7\) random permutations of the
daily values, are \(9.98\times10^{-4}\) and \(8.95\times10^{-4}\).
These correspond to equivalent Gaussian significances of
\(3.09\sigma\) and \(3.12\sigma\), respectively. As a
supplementary check, the same analysis was repeated using daily
median DOL, yielding similar peaks near 120 days. Folding the
daily mean series at 119.91 days shows a phase-dependent
variation. A descriptive sinusoidal fit to all 95 daily means has
a semi-amplitude of 3.23\%. The time-aligned DOL
and central-frequency fits show local similarities, with peak
offsets that change across the observing span. Geometric and
environmental modulation should be involved in the approximately
120-day DOL variation.

\section*{Acknowledgments}
This work is supported by the National Natural Science Foundation of China (Grant Nos. 12233002, 12273113, and 12403046),
by the China Postdoctoral Science Foundation (Grant No. 2025M783225),
and by the Specialized Research Fund for State Key Laboratory of Radio Astronomy and Technology (No. SDQZ-KF-A2610).

\bibliographystyle{aasjournal}
\bibliography{ref}

\clearpage

\end{document}